\documentclass[preprintnumbers,article,amsmath,amssymb,floatfix,10pt,prd,twocolumn,
superscriptaddress,nofootinbib]{revtex4-2}
\usepackage{bm}
\usepackage{amsfonts}
\usepackage{latexsym}
\usepackage[latin1]{inputenc}
\usepackage{graphicx}
\usepackage{amsmath}
\usepackage{palatino}
\usepackage{mathpazo}
\usepackage{textcomp}
\usepackage{float}
\usepackage{booktabs}
\usepackage{dcolumn}
\usepackage{ragged2e}
\usepackage{hyperref}
\hypersetup{colorlinks,citecolor=blue}
\hypersetup{colorlinks=true,linkcolor=red,filecolor=magenta,    urlcolor=blue}
\usepackage{amsmath}
\usepackage{xcolor}
\usepackage{orcidlink}
\usepackage{epsfig}
\usepackage{caption}
\usepackage{subcaption}
\usepackage{commath}
\def\jnl@style{\it}
\def\aaref@jnl#1{{\jnl@style#1}}

\def\aaref@jnl#1{{\jnl@style#1}}

\def\aj{\aaref@jnl{AJ}}                   
\def\apj{\aaref@jnl{ApJ}}                 
\def\apjl{\aaref@jnl{ApJ}}                
\def\apjs{\aaref@jnl{ApJS}}               
\def\apss{\aaref@jnl{Ap\&SS}}             
\def\aap{\aaref@jnl{A\&A}}                
\def\aapr{\aaref@jnl{A\&A~Rev.}}          
\def\aaps{\aaref@jnl{A\&AS}}              
\def\mnras{\aaref@jnl{Mon.~Not.~Roy.~Astron.~Soc.}}             
\def\prd{\aaref@jnl{Phys.~Rev.~D}}        
\def\prc{\aaref@jnl{Phys.~Rev.~C}}  
\def\prl{\aaref@jnl{Phys.~Rev.~Lett.}}    
\def\qjras{\aaref@jnl{QJRAS}}             
\def\skytel{\aaref@jnl{S\&T}}             
\def\ssr{\aaref@jnl{Space~Sci.~Rev.}}     
\def\zap{\aaref@jnl{ZAp}}                 
\def\nat{\aaref@jnl{Nature}}              
\def\aplett{\aaref@jnl{Astrophys.~Lett.}} 
\def\apspr{\aaref@jnl{Astrophys.~Space~Phys.~Res.}} 
\def\physrep{\aaref@jnl{Phys.~Rep.}}      
\def\physscr{\aaref@jnl{Phys.~Scr}}       
\def\commat{\aaref@jnl{Comm.~Math.~Phys.}}              
\def\science{\aaref@jnl{Science}}               
\def\cqg{\aaref@jnl{Classical Quant.~Grav.}}            
\def\jpcs{\aaref@jnl{JPCS}}                                     
\def\ijmpd{\aaref@jnl{Int.~J.~Mod.~Phys.~D}}                    
\def\grg{\aaref@jnl{Gen.~Relat.~Gravit.}}               
\def\rpp{\aaref@jnl{Rep.~Prog.~Phys.}}          
\def\npa{\aaref@jnl{Nucl.~Phys.~A}}        
\def\lrr{\aaref@jnl{Living Rev.~Rel.}}                   
\def\jcap{\aaref@jnl{J.~Cosmology Astropart.~Phys.}}    
\def\rmp{\aaref@jnl{Rev.~Mod.~Phys.}}   
\def\epjc{\aaref@jnl{Eur.~Phys.~J.~C}} 
\def\plb{\aaref@jnl{~Phy.~Lett.~B}} 
\def\mpla{\aaref@jnl{Mod.~Phy.~Lett.~A}} 
\def\arxiv{\aaref@jnl{arxiv.org}}

\allowdisplaybreaks[1]

\begin{document}
\color{black}       
\title{Probing Weyl-type $f(Q, T)$ gravity: Cosmological implications and constraints}

\author{Alnadhief H. A. Alfedeel\orcidlink{0000-0002-8036-268X}}%
\email[Email: ]{aaalnadhief@imamu.edu.sa}
\affiliation{Department of Mathematics and Statistics, Imam Mohammad Ibn Saud Islamic University (IMSIU),\\
Riyadh 13318, Saudi Arabia.}
\affiliation{Department of Physics, Faculty of Science, University of Khartoum, P.O. Box 321, Khartoum 11115, Sudan.}
\affiliation{Centre for Space Research, North-West University, Potchefstroom 2520, South Africa.}

\author{M. Koussour\orcidlink{0000-0002-4188-0572}}
\email[Email: ]{pr.mouhssine@gmail.com}
\affiliation{Department of Physics, University of Hassan II Casablanca, Morocco.} 

\author{N. Myrzakulov\orcidlink{0000-0001-8691-9939}}
\email[Email: ]{nmyrzakulov@gmail.com}
\affiliation{L. N. Gumilyov Eurasian National University, Astana 010008,
Kazakhstan.}

%

\begin{abstract}
In this paper, we investigate the cosmological implications and constraints of Weyl-type $f(Q, T)$ gravity. This theory introduces a coupling between the non-metricity $Q$ and the trace $T$ of the energy-momentum tensor, using the principles of proper Weyl geometry. In this geometry, the scalar non-metricity $Q$, which characterizes the deviations from Riemannian geometry, is expressed in its standard Weyl form $\nabla _{\mu }g_{\alpha \beta
}=-w_{\mu }g_{\alpha \beta }$ and is determined by a vector field $w_{\mu }$. To study the implications of this theory, we propose a deceleration parameter with a single unknown parameter $\chi$, which we constrain by using the latest cosmological data. By solving the field equations derived from Weyl-type $f(Q, T)$ gravity, we aim to understand the behavior of the energy conditions within this framework. In the present work, we consider two well-motivated forms of the function $f(Q, T)$: (i) the linear model represented by $f(Q, T) = \alpha Q + \frac{\beta}{6\kappa^2} T$, and (ii) the coupling model represented by $f(Q, T) = \frac{\gamma}{6H_0^2 \kappa^2} QT$, where $\alpha$, $\beta$, and $\gamma$ are free parameters. Here, $\kappa^2 = \frac{1}{16\pi G}$ represents the gravitational coupling constant. In both of the models considered, the strong energy condition is violated, indicating consistency with the present accelerated expansion. However, the null, weak, and dominant energy conditions are satisfied in these models.

\end{abstract}
\date{\today}
\maketitle

\section{Introduction}
\label{sec1}

Recent astrophysical observations from various sources, such as type Ia supernovas \cite{Riess,Perlmutter}, cosmic microwave background anisotropies \cite{R.R.,Z.Y.}, large-scale structures \cite{T.Koivisto,S.F.} and baryon acoustic oscillations \cite{D.J.,W.J.} have provided strong evidence for the accelerating expansion of the Universe in the present epoch. This intriguing phenomenon is attributed to the dominance of a mysterious energy component, known as Dark Energy (DE), which possesses a large negative pressure. While General Relativity (GR) is the standard theory of gravity, it has not yet provided a satisfactory explanation for the nature and origin of DE.

To address this issue, numerous alternative models have been proposed within the framework of modified theories of gravity. One such approach is $f(R)$ gravity, where the traditional Einstein-Hilbert action is modified by replacing the curvature scalar $R$ with an arbitrary function of $R$ \cite{R1, R2, R3}. Another alternative theory is $f(\mathcal{T})$ gravity, which considers the torsion scalar $\mathcal{T}$ in the teleparallel approach \cite{T1, T2, T3, T4, T5}. In addition, Gauss-Bonnet $f(G)$ gravity \cite{G1} and $f(Q)$ gravity \cite{Q1} are among the other alternative theories of gravity that have been explored. The motivation behind these modified theories of gravity lies in the search for a suitable candidate that can account for the observed cosmic acceleration and provide a deeper understanding of the nature of DE.

The teleparallel approach to gravity offers an alternative perspective on describing the gravitational properties of spacetime. In this approach, the metric tensor $g_{\alpha\beta}$, which is traditionally used in GR, is replaced by a set of tetrad vectors $e_{\alpha}^{i}$. These tetrad fields give rise to torsion, which can fully account for gravitational effects, replacing the need for curvature. This formulation leads to the development of the Teleparallel Equivalent of GR (TEGR), also referred to as $f(\mathcal{T})$ gravity theory. The TEGR theory was initially introduced in \cite{T6,T7}. and has since gained recognition as a viable framework for understanding gravity. In $f(\mathcal{T})$ gravity, the torsion plays a central role, providing a new perspective on the nature of gravitational interactions. In teleparallel or $f(\mathcal{T})$ type theories, the presence of torsion precisely cancels out the effects of curvature, resulting in a flat spacetime. This intriguing feature distinguishes these theories from others and highlights their unique characteristics. One notable advantage of the $f(\mathcal{T})$ gravity theory is that its field equations are second-order differential equations.

Recently, a novel approach has emerged, shifting the focus of gravitational interactions to the non-metricity $Q$ of the metric \cite{Q1}. Furthermore, the exploration of the non-metricity $Q$ of the metric has led to the development of a new geometry known as Weyl's geometry. In Weyl's geometry, the gravitational effects arise not from the rotation of the angle between two vectors during parallel transport, but rather from the variation in the length of the vector itself, which mathematically describes as $Q_{\mu \alpha \beta }=\nabla _{\mu }g_{\alpha \beta }\neq0$. This unique perspective shifts the focus to the intrinsic properties of the vectors, revealing that changes in vector length play a fundamental role in describing gravitational phenomena. This innovative gravitational framework, known as the Symmetric Teleparallel Equivalent of GR (STEGR) gravity, was first introduced in \cite{Q2}. Over time, it has evolved into the $f(Q)$ gravity theory, also referred to as coincident GR or nonmetric gravity, as documented in \cite{Q1}. In the last two decades, extensive research has been conducted to explore the geometric and physical aspects of STEGR gravity, and recent years have witnessed a surge in interest in this theory \cite{Q3,Q4,Q5,Q6,Q7,Q8}. For a comprehensive overview of teleparallel gravity, refer to Ref. \cite{Q9}.

The $f(Q,T)$ gravity represents a generalization of $f(Q)$ gravity, where the trace of the energy-momentum tensor $T$ is taken into account \cite{QT1}. This extension allows for the incorporation of additional effects arising from exotic imperfect fluids or quantum phenomena, which can introduce dependence on $T$. The cosmological implications and reconstruction of $f(Q,T)$ gravity have been extensively investigated in recent literature, aiming to understand the impact of these modifications on the dynamics of the Universe and to explore their compatibility with observational data. In this study \cite{QT2}, the authors investigate the effects of perturbations on the cosmological dynamics within the context of $f(Q, T)$ gravity. They analyze the behavior of linear perturbations in the early and late Universe, aiming to understand the impact of $f(Q, T)$ gravity on the evolution of cosmological structures and the generation of observable features. The authors of \cite{QT3,QT4} explore the behavior of inflationary models in the context of $f(Q, T)$ gravity and examine their compatibility with observational constraints. Narawade et al. \cite{QT5} investigated the cosmological implications of this theory by constructing a specific accelerating cosmological model. Using the dynamical system analysis technique, the authors study the evolution of the cosmological model and analyze its stability and attractor behavior. They explore the parameter space of the model and identify the regions that lead to accelerated expansion, consistent with the observational data on the accelerating Universe (please see also \cite{QT6}).

In this study, we investigate a specific formulation of the $f(Q, T)$ gravity theory, focusing on the non-minimal coupling between the non-metricity $Q$ and the trace $T$ of the matter energy-momentum tensor. Our analysis is conducted within the framework of the proper Weyl geometry, where we adopt a specific expression for the non-metricity $Q$ derived from the non-conservation of the divergence of the metric tensor, $\nabla _{\mu }g_{\alpha \beta
}=-w_{\mu }g_{\alpha \beta }$. By using this approach, we are able to describe the non-metricity in terms of a vector field $w_{\mu }$, in conjunction with the metric tensor. In other words, the non-metricity is fully determined by the magnitude of the vector field $w_{\mu }$. This formulation is known as Weyl-type $f(Q, T)$ gravity \cite{Weyl1}. Through this approach, we aim to explore the constraints and implications of the energy conditions in Weyl-type $f(Q, T)$ gravity theory, with its specific coupling between the non-metricity and trace of the matter energy-momentum tensor. Yang et al. \cite{Weyl2} derived the Newtonian and post-Newtonian limits of Weyl-type $f(Q, T)$ gravity. This analysis helps in obtaining constraints imposed by Solar System-level gravity on the theory and on the properties of the Weyl vector. Additionally, constraints from other astrophysical observations can also be derived using the Newtonian limit. Also, the authors investigated various aspects of Weyl-type $f(Q, T)$ gravity, including geodesic deviation, the Raychaudhuri equation, and tidal forces within this framework. Koussour \cite{Weyl3} introduces a model-independent approach within the framework of Weyl-type $f(Q, T)$ gravity to study the crossing of the phantom divide line. By considering the non-metricity $Q$ and the trace $T$ of the matter energy-momentum tensor, the author investigates the behavior of the phantom divide-line in a linear cosmological model. In this study, our main objective is to investigate the behavior of various energy conditions within the framework of the Weyl-type $f(Q, T)$ gravity theory. Energy conditions hold great significance in cosmology, black hole thermodynamics \cite{ECs1}, and singularity theorems \cite{ECs2} within the realm of GR. They provide different criteria for ensuring the positivity of the energy-momentum tensor in the presence of matter and the attractive nature of gravity. These energy conditions are derived from the Raychaudhuri equation \cite{Raychaudhuri}, which exhibits their purely geometric nature and requires the energy density to be positive in order to maintain the attractive nature of gravity. In their comprehensive work, Capozziello et al. \cite{ECs3} extensively examined the concept of generalized energy conditions in extended theories of gravity. They provided a detailed description of these energy conditions, which involved considering the contraction of timelike and null vectors in relation to various tensors, such as the energy-momentum tensor, Ricci tensor, and Einstein tensor. Furthermore, Capozziello et al.\cite{ECs4,ECs5} contributed to this understanding by offering clear explanations of energy conditions within the context of modified gravities.

The structure of this study is as follows: In Sec. \ref{sec2}, we provide an introduction to the fundamental concepts of Weyl-type $f(Q,T)$ gravity. Sec. \ref{sec3} is dedicated to the modified Friedmann equations in Weyl-type $f(Q,T)$ gravity, where we also present the two well-motivated forms of the function $f(Q, T)$: (i) the linear model represented by $f(Q, T) = \alpha Q + \frac{\beta}{6\kappa^2} T$, and (ii) the coupling model represented by $f(Q, T) = \frac{\gamma}{6H_0^2 \kappa^2} QT$. The discussion on energy conditions is presented in Sec. \ref{sec4}. In Sec. \ref{sec5}, to study the implications of this theory, we propose a deceleration parameter with a single unknown parameter $\chi$ and apply it to the modified Friedmann equations of Weyl-type $f(Q,T)$ gravity. Finally, in the concluding Secs. \ref{sec6} and \ref{sec7}, we present the obtained results and provide a comprehensive summary.

\section{Introduction of Weyl-type $f(Q,T)$ gravity: A comprehensive overview}
\label{sec2}

The action governing Weyl-type $f(Q,T)$ gravity can be expressed as \cite{Weyl1,Weyl2}, 
\begin{multline}
S=\int \sqrt{-g}d^{4}x \left[ \kappa ^{2}f(Q,T)-\frac{1}{4}W_{\alpha \beta }W^{\alpha \beta
}-\frac{1}{2}m^{2}w_{\alpha }w^{\alpha }+\right.   \label{1} \\
\left. \lambda (R+6\nabla _{\mu }w^{\mu }-6w_{\mu }w^{\mu })+\mathcal{L}_{m}%
\right].
\end{multline}

In this context, the field strength tensor of the vector field is denoted as $W_{\alpha \beta }=\nabla _{\beta }w_{\alpha }-\nabla _{\alpha
}w_{\beta \text{ }}$, where $w_{\alpha}$ represents the vector field itself. The quantity $\kappa ^{2}=1/16\pi G$, with $G$ being the gravitational constant, and $m$ denotes the particle mass associated with the vector field. The matter Lagrangian is denoted as $\mathcal{L}_{m}$. The action comprises of three terms. The first term corresponds to the gravitational interaction described by the Weyl-type $f(Q,T)$ function. The second term represents the ordinary kinetic term of the vector field, while the third term accounts for the mass term associated with the vector field. It should be noted that the function $f(Q,T)$ denotes an arbitrary function of the non-metricity scalar $Q$ and the trace of the matter-energy-momentum tensor $T$. In addition, the symbol $\lambda$ represents the Lagrange multiplier. It is introduced as a parameter to enforce constraints or conditions within the theory, especially, imposing the flat geometry constraint i.e. the total curvature vanishing of the Weyl space ($\bar{R}=0$).

The scalar of the non-metricity, denoted as $Q$, assumes a crucial role within our theory, significantly influencing its dynamics. It is defined as, 
\begin{equation}
Q\equiv -g^{\alpha \beta }\left( L_{\nu \beta }^{\mu }L_{\beta \mu }^{\nu
}-L_{\nu \mu }^{\mu }L_{\alpha \beta }^{\nu }\right) ,  \label{2}
\end{equation}%
where $L_{\alpha \beta }^{\lambda }$ denotes the tensor of deformation read as, 
\begin{equation}
L_{\alpha \beta }^{\lambda }=-\frac{1}{2}g^{\lambda \gamma }\left( Q_{\alpha
\gamma \beta }+Q_{\beta \gamma \alpha }-Q_{\gamma \alpha \beta }\right) .
\label{3}
\end{equation}

In Riemannian geometry, both the Levi-Civita connection $\Gamma _{\alpha
\beta }^{\lambda }$ and the metric tensor $g_{\alpha\beta}$ can be compatible, meaning that their covariant derivative with respect to the connection vanishes, i.e., $\nabla_{\mu} g_{\alpha\beta} = 0$. However, in Weyl's geometry, this compatibility seems to be altered. The non-metricity captures the deviation from metric compatibility, providing insights into the geometric properties of the spacetime under consideration. So, we have,
\begin{equation}
\overline{Q}_{\mu \alpha \beta }\equiv \overline{\nabla }_{\mu }g_{\alpha
\beta }=\partial _{\mu }g_{\alpha \beta }-\overline{\Gamma }_{\mu \alpha
}^{\rho }g_{\rho \beta }-\overline{\Gamma }_{\mu \beta }^{\rho }g_{\rho
\alpha }=2w_{\mu }g_{\alpha \beta },  \label{4}
\end{equation}%
where, 
\begin{equation}
    \overline{\Gamma }_{\alpha \beta }^{\lambda }\equiv \Gamma _{\alpha
\beta }^{\lambda }+g_{\alpha \beta }w^{\lambda }-\delta _{\alpha }^{\lambda
}w_{\beta }-\delta _{\beta }^{\lambda }w_{\alpha },
\end{equation}
and $\Gamma _{\alpha
\beta }^{\lambda }$ represent the semi-metric connection in Weyl geometry and the Christoffel symbol in terms of the metric
tensor $g_{\alpha \beta }$, respectively. Weyl introduced the semi-metric connection to capture the joint variation of both direction and magnitude experienced by a vector field.

Taking into account Eqs. \eqref{2}-\eqref{4}, we are able to derive the following relationship:
\begin{equation}
Q=-6w^{2}.  \label{5}
\end{equation}

By performing the variation of the action with respect to the vector field, we derive the generalized Proca equation, which governs the evolution of the field,
\begin{equation}\label{EOM1}
    \nabla^\beta W_{\alpha \beta }-(m^2+12\kappa^2 f_Q+12\lambda)w_\alpha=6\nabla_\alpha \lambda.
\end{equation}

Upon comparing Eq. (\ref{EOM1}) with the standard Proca equation, we observe that the effective dynamical mass of the vector field can be expressed as follows:
\begin{equation}\label{discussion1}
m^2_{\rm{eff}}=m^2+12\kappa^2f_Q+12\lambda.
\end{equation}
where $f_{Q}\equiv \frac{%
\partial f(Q,T)}{\partial Q}$. It is noteworthy that the Lagrange multiplier field yields an effective current for the vector field. In the realm of quantum field theory, experimental measurements often reveal deviations between the observed mass and the bare mass, which can be attributed to the presence of interactions. So, Eq. (\ref{discussion1}) highlights that within the framework of Weyl-type $f(Q,T)$ gravity, such deviations in mass can also stem from the nontrivial geometric characteristics of the spacetime.

Moreover, the generalized field equation is obtained by varying the action with respect to the metric tensor, as governed by Eq. \eqref{1}, 
\begin{multline}
\frac{1}{2}\left( T_{\alpha \beta }+S_{\alpha \beta }\right) -\kappa
^{2}f_{T}\left( T_{\alpha \beta }+\Theta _{\alpha \beta }\right) =-\frac{%
\kappa ^{2}}{2}g_{\alpha \beta }f(Q,T)  \label{7} \\
-6\kappa ^{2}f_{Q}w_{\alpha }w_{\beta }+\lambda \left( R_{\alpha \beta
}-6w_{\alpha }w_{\beta }+3g_{\alpha \beta }\nabla _{\rho }w^{\rho }\right) 
\\
+3g_{\alpha \beta }w^{\rho }\nabla _{\rho }\lambda -6w_{(\alpha }\nabla
_{\beta )}\lambda +g_{\alpha \beta }\square \lambda -\nabla _{\alpha }\nabla
_{\beta }\lambda ,
\end{multline}%
where 
\begin{equation}
T_{\alpha \beta }\equiv -\frac{2}{\sqrt{-g}}\frac{\delta (\sqrt{-g}\mathcal{L}_{m})}{%
\delta g^{\alpha \beta }},  \label{8}
\end{equation}%
and%
\begin{equation}
f_{T}\equiv \frac{\partial f(Q,T)}{\partial T},  \label{9}
\end{equation}%
respectively. Furthermore, we define the quantity $\Theta_{\alpha\beta}$ as follows: 
\begin{equation}
\Theta _{\alpha \beta }\equiv g^{\mu \nu }\frac{\delta T_{\mu \nu }}{\delta
g_{\alpha \beta }}=g_{\alpha \beta }\mathcal{L}_{m}-2T_{\alpha \beta }-2g^{\mu \nu }%
\frac{\delta ^{2}\mathcal{L}_{m}}{\delta g^{\alpha \beta }\delta g^{\mu \nu }}.
\label{10}
\end{equation}

Within the aforementioned field equation, $S_{\alpha\beta}$ represents the rescaled energy-momentum tensor associated with the free Proca field,
\begin{equation}
S_{\alpha \beta }=-\frac{1}{4}g_{\alpha \beta }W_{\rho \sigma }W^{\rho
\sigma }+W_{\alpha \rho }W_{\beta }^{\rho }-\frac{1}{2}m^{2}g_{\alpha \beta
}w_{\rho }w^{\rho }+m^{2}w_{\alpha }w_{\beta }.  \label{11}
\end{equation}

In the Weyl-type $f(Q,T)$ theory, the divergence of the matter-energy-momentum tensor can be expressed as follows \cite{Weyl1}:
\begin{align}\label{div}
    \nabla^\alpha T_{\alpha\beta}=\frac{\kappa^2}{1+2\kappa^2 f_T}\Big [ 2\nabla_\beta(\mathcal{L}_{m}f_T)-f_T\nabla_\beta T-2T_{\alpha\beta}\nabla^\alpha f_T\Big ].
\end{align}

Thus, the equation presented above demonstrates that in the Weyl-type $f(Q,T)$ theory, the matter energy-momentum tensor does not exhibit conservation. Also, it is important to highlight that when $f_T = 0$, the energy-momentum tensor becomes conserved.

\section{Cosmological Weyl-type $f(Q,T)$ models}

\label{sec3}

Let us assume that the Universe is described by a homogeneous, isotropic, and spatially Friedmann-Lemaitre-Robertson-Walker (FLRW) line element. This line element, which captures the overall geometry of the Universe, can be expressed as, 
\begin{equation}
ds^2 = -dt^2 + a^2(t) \left( \frac{dr^2}{1 - kr^2} + r^2 d\Omega^2 \right),  \label{FLRW}
\end{equation}%
where $t$ represents the cosmic time, $a(t)$ denotes the scale factor representing the expansion of the Universe, $r$ is the comoving radial coordinate, $k$ corresponds to the curvature of the spatial sections (taking values $k = -1, 0, 1$ for open, flat, and closed Universes, respectively), and $d\Omega^2$ represents the line element of the unit 2-sphere. In this study, we specifically focus on the flat case of the FLRW model i.e. $k =0$. In addition, we assume that the vector field can be characterized as,
\begin{equation}
w_{\alpha }=\left[
\psi (t),0,0,0\right].
\end{equation}

Thus, $w^{2}=w_{\alpha }w^{\alpha }=-\psi
^{2}(t)$ and $Q=-6w^{2}=6\psi ^{2}(t)$.

Furthermore, we assume that the Universe can be described as a perfect fluid in which the energy-momentum tensor is defined as,
\begin{equation}
\label{EMT}
T_{\mu\nu}=\left(\rho+p\right)u_\mu u_\nu+ p g_{\mu\nu},
\end{equation}
where $\rho$ represents the energy density, $p$ denotes the pressure, and $u^\alpha$ corresponds to the 4-velocity of the fluid ($u_\alpha u^\alpha=-1$). This suggests that $T^\alpha_\beta=diag\left(-\rho,p,p,p\right)$, and $\Theta^\alpha_\beta=\delta^\alpha_\beta p-2T^\alpha_\beta=diag\left(2\rho+p,-p,-p,-p\right)$.

In the cosmological scenario, the constraints of flat space and the generalized Proca equation can be expressed as, 
\begin{eqnarray}
\dot{\psi} &=&\dot{H}+2H^{2}+\psi ^{2}-3H\psi ,  \label{17} \\
\dot{\lambda} &=&\left( -\frac{1}{6}m^{2}-2\kappa ^{2}f_{Q}-2\lambda \right)
\psi =-\frac{1}{6}m_{eff}^{2}\psi ,  \label{18} \\
\partial _{i}\lambda &=&0.  \label{19}
\end{eqnarray}
where $H(t) = \frac{\dot{a}}{a}$ represents the Hubble parameter, which characterizes the rate of expansion of the Universe, and dot (.) represents the derivative
with respect to time t.

Using Eq. (\ref{7}) and employing the provided metric (\ref{FLRW}), we derive the generalized Friedmann equations as \cite{Weyl1}, 
\begin{align}
\kappa ^{2}f_{T}(\rho +p)& +\frac{1}{2}\rho =\frac{\kappa ^{2}}{2}f-\left(
6\kappa ^{2}f_{Q}+\frac{1}{4}m^{2}\right) \psi ^{2}  \notag \\
& -3\lambda (\psi ^{2}-H^{2})-3\dot{\lambda}(\psi -H),  \label{F1}
\end{align}%
\begin{align}
-\frac{1}{2}p& =\frac{\kappa ^{2}}{2}f+\frac{m^{2}\psi ^{2}}{4}+\lambda
(3\psi ^{2}+3H^{2}+2\dot{H})  \notag \\
& +(3\psi +2H)\dot{\lambda}+\ddot{\lambda}.  \label{F2}
\end{align}

The generalized Friedmann equations (\ref{F1}) and (\ref{F2}) can be equivalently expressed in an effective form as,
\begin{equation}
\label{gen1}
3H^2=\frac{1 }{2\lambda}\left(\rho+\rho _{eff}\right),
\end{equation}
\begin{equation}
    2\dot{H}=-\frac{1}{2\lambda }\left(\rho +\rho _{eff}+p+p_{eff}\right)
\end{equation}
where
\begin{equation}
\label{rhoeff}
\rho _{eff}=m_{{\rm eff}}^2H\psi+2\kappa ^2f_T\left(\rho +p\right)-\kappa ^2f-\frac{m^2\psi ^2}{2}-6\lambda \psi ^2,
\end{equation}
and
\begin{eqnarray}
\label{peff}
p_{eff}&=&\frac{m_{{\rm eff}}^2}{3}\left(\dot{\psi}+\psi ^2-4H\psi\right)+\kappa ^2f+4\kappa ^2\dot{f}_Q\psi \nonumber\\
&&+\frac{m^2\psi ^2}{2}+6\lambda \psi ^2,
\end{eqnarray}
respectively. In the special case where $f=0$, $\psi=0$, and $\lambda=\kappa^2$, the gravitational action (\ref{1}) simplifies to the standard Hilbert-Einstein action. As a consequence, the generalized equations (\ref{F1}) and (\ref{F2}) reduce to the standard Friedmann equations in GR. Specifically, these equations become $3H^2=\frac{\rho}{2\kappa^2}$ and $2\dot{H}=-\frac{(\rho+p)}{2\kappa^2}$, respectively. In this limit, the energy density $\rho$ and pressure $p$ behave according to the standard framework of GR.

In order to examine the behavior of energy conditions, we will examine various cosmological models within the framework of Weyl-type $f(Q, T)$ gravity theory. These models correspond to different selections of the function $f(Q, T)$, which characterizes the non-minimal coupling between the scalar non-metricity and matter. Additionally, we will assume an additional equation to study the dynamics of geometric and physical cosmological quantities in Weyl-type $f(Q, T)$ gravity. We consider two specific functional forms for $f(Q, T)$ as: $f(Q,T)=\alpha Q+\frac{\beta }{6\kappa ^2}T$ (linear model) and $f(Q,T)=\frac{\gamma}{6H_0^2\kappa ^2}QT$ (coupling model). In addition, we assume that $M^2= m^{2}/\kappa^2$, which represents the dimensionless quantity related to the mass of the Weyl vector field, indicating the strength of the coupling between Weyl geometry and matter. In this work, we consider a value of $M= 0.95$ for the mass parameter, with $\lambda=\kappa^2=1$ \cite{Weyl1}.

\subsection{$f(Q,T)=\alpha Q+\frac{\beta }{6\kappa ^2}T$}

For the first cosmological model in Weyl-type $f(Q, T)$ gravity, we will examine the scenario where the function $f(Q, T)$ can be expressed as $f(Q,T)=\alpha Q+\frac{\beta }{6\kappa ^2}T$, where $\alpha$ and $\beta$ are constants. Hence, $f_{Q}=\alpha$ and $f_{T}=\frac{\beta }{6\kappa ^2}$. By selecting this specific form for the function $f(Q, T)$ and solving Eqs. (\ref{F1}) and (\ref{F2}), we derive the expressions for the pressure $p$, energy density $\rho$, and equation of state (EoS) parameter $\omega=\frac{p}{\rho}$ as,
\begin{widetext}
\begin{eqnarray}
\label{rho1}
    \rho&=&-\frac{3 \left[H^2 \left(12 (2 \alpha  \beta +3 \alpha +\beta )+(2 \beta +3) M^2\right)+4 \beta  \dot{H}\right]}{2 \left(2 \beta ^2+9 \beta +9\right)}, \\
    p&=&-\frac{3 \left[12 (\beta +2) \dot{H}+H^2 \left(12 (2 \alpha  \beta +3 \alpha +3 \beta +6)+(2 \beta +3) M^2\right)\right]}{2 \left(2 \beta ^2+9 \beta +9\right)}, \\
\label{p1}
    \omega&=&\frac{H^2 \left[24 \alpha  \beta +36 (\alpha +\beta +2)+(2 \beta +3) M^2\right]+12 (\beta +2) \dot{H}}{H^2 \left(12 (2 \alpha  \beta +3 \alpha +\beta )+(2 \beta +3) M^2\right)+4 \beta  \dot{H}}.
\label{omega1}
\end{eqnarray}
\end{widetext}

\subsection{$f(Q,T)=\frac{\gamma}{6H_0^2\kappa ^2}QT$}

For the second cosmological model in Weyl-type $f(Q, T)$ gravity, we will examine the scenario where the function $f(Q, T)$ can be expressed as $f(Q,T)=\frac{\gamma}{6H_0^2\kappa ^2}QT$, where $\gamma$ is a constant. Hence, $f_{Q}=\frac{\gamma  T}{6 H_0^2 \kappa ^2}$ and $f_{T}=\frac{\gamma  Q}{6 H_0^2 \kappa ^2}$. By selecting this specific form for the function $f(Q, T)$ and solving Eqs. (\ref{F1}) and (\ref{F2}), we derive the expressions for the pressure $p$, energy density $\rho$, and EoS parameter $\omega$ as,
\begin{widetext}
\begin{eqnarray}
\label{p2}
    \rho&=&\frac{H^2 H_{0}^2 \left[2 \gamma  H^2 \left(M^2+60\right)+40 \gamma  \dot{H}-H_{0}^2 M^2\right]}{2 \left(8 \gamma ^2 H^4+4 \gamma  H^2 H_{0}^2+H_{0}^4\right)}, \\
    p&=&-\frac{H_{0}^2 \left[2 \gamma  H^4 \left(M^2+12\right)+8 \dot{H} \left(\gamma  H^2+H_{0}^2\right)+H^2 H_{0}^2 \left(M^2+24\right)\right]}{2 \left(8 \gamma ^2 H^4+4 \gamma  H^2 H_{0}^2+H_{0}^4\right)}, \\
    \label{rho2}
    \omega&=&-\frac{2 \gamma  H^4 \left(M^2+12\right)+8 \dot{H} \left(\gamma  H^2+H_{0}^2\right)+H^2 H_{0}^2 \left(M^2+24\right)}{H^2 \left[2 \gamma  H^2 \left(M^2+60\right)+40 \gamma  \dot{H}-H_{0}^2 M^2\right]}.
    \label{omega2}
\end{eqnarray}
\end{widetext}

\section{Energy conditions}

\label{sec4}

In this section, we will examine the energy conditions in specific models of Weyl-type $f(Q,T)$ gravity. By considering different choices of the function $f(Q,T)$, we will explore how these models satisfy or deviate from the energy conditions commonly used in GR. In the study of the Universe, certain physical parameters such as the deceleration parameter and the EoS parameter play a crucial role. However, an important aspect of modern cosmology involves investigating energy conditions, which are derived from the Raychaudhuri equation \cite{Raychaudhuri}. The Raychaudhuri equation is the foundation for establishing energy conditions that satisfy the attractive nature of gravity and ensure the positivity of energy density. These conditions are essential for maintaining the consistency of null, timelike, and lightlike geodesics. In Riemann geometry, the Raychaudhuri equation is expressed as \cite{Santos},
\begin{equation}  \label{16}
\frac{d\theta}{d\tau}=-\frac{1}{3}\theta^2-\sigma_{\mu\nu}\sigma^{\mu\nu}+%
\omega_{\mu\nu}\omega^{\mu\nu}-R_{\mu\nu}u^{\mu}u^{\nu}\,,
\end{equation}
where $R_{\mu\nu}$ is the Ricci tensor, $\theta$ is the expansion factor, and $\sigma^{\mu\nu}$ and $\omega_{\mu\nu}$ correspond to the shear
and rotation associated with the vector field $u^{\mu}$. For gravity
exhibiting attractive behavior, Eq. \eqref{16} fulfill the following condition: 
\begin{equation}
 \label{18}
R_{\mu\nu}u^{\mu}u^{\nu}\geq0\,.
\end{equation}

In the context of Weyl geometry with non-metricity, the Raychaudhuri equation is modified to account for the specific geometric properties as \cite{Weyl2},
\begin{widetext}
\begin{eqnarray}
\left[ \theta -2\omega _{\mu }u^{\mu }\right] ^{^{\prime }} &=&-\frac{1}{3}%
\left[ \theta -\frac{\left( l^{2}\right) ^{^{\prime }}}{2l^{2}}-\omega _{\mu
}u^{\mu }\right] ^{2}+\left[ \frac{\left( l^{2}\right) ^{^{\prime }}}{2l^{2}}%
-\omega _{\mu }u^{\mu }\right] ^{2}-R_{\mu \nu }u^{\mu }u^{\nu }-\sigma
_{\mu \nu }\sigma ^{\mu \nu }+\omega _{\mu \nu }\omega ^{\mu \nu }+\nabla
^{\mu }f_{\mu }  \nonumber \\
&&-\frac{1}{l^{2}}f^{\mu }\nabla _{\mu }l^{2}-2\omega _{\mu }f^{\mu
}-2\left( \omega _{\mu }u^{\mu }\right) ^{2}-2u^{\mu }u^{\nu }\nabla _{\mu
}\omega _{\nu }+l^{2}\nabla ^{\mu }\omega _{\mu }-2l^{2}\omega _{\mu }\omega
^{\mu },
\label{Ray}
\end{eqnarray}
\end{widetext}
where the equation describes the specific form of the extra force in Weyl-type $f (Q, T)$ gravity given by
\begin{equation}
    f^{\rho }=\frac{l^{2}}{p+\rho }h^{\rho \nu }\widetilde{\nabla }^{\mu }\left(
T_{\mu \nu }-pg_{\mu \nu }\right) +\frac{\left( l^{2}\right) ^{^{\prime }}}{%
2l^{2}}u^{\rho }-\omega _{\mu }u^{\mu }u^{\rho }.
\label{extra}
\end{equation}

Here, $l(x^\mu)$ is an arbitrary function of space and time coordinates, $h^{\alpha\beta}$ is a generalized projection tensor operator in Weyl geometry \cite{Weyl2}. It is important to note that the influence of the non-minimal coupling between matter and geometry, represented by $f(Q, T)$, affects the Raychaudhuri equation (\ref{Ray}) through the expression of an extra force as shown in Eq. (\ref{extra}). In addition, when $f_T = 0$, corresponding to the minimal coupling between matter and geometry, the above equation simplifies to the generalized Raychaudhuri equation in the coincidence gravity theory \cite{Q1}. The coincidence gravity theory is a broader framework than symmetric teleparallel gravity. Hence, considering Eq. (\ref{Ray}) with $f_T = 0$ can be regarded as the Raychaudhuri equation of the generalized STEGR.

The energy conditions are employed to investigate the expansion of the Universe. Several forms of energy conditions exist, including the null
energy condition (NEC), weak energy condition (WEC), dominant energy
condition (DEC), and strong energy condition (SEC). Thus, when considering a
perfect fluid matter distribution, the energy conditions are defined as,
\begin{itemize}
\item NEC if $\rho_{eff}+p_{eff}\geq 0\,$;

\item WEC if $\rho_{eff}\geq 0, \rho_{eff}+p_{eff}\geq 0\,$;

\item DEC if $\rho_{eff}\geq 0, |p_{eff}|\leq \rho_{eff}\,$;

\item SEC if $\rho_{eff}+3\,p_{eff}\geq 0$.
\end{itemize}

Here, $\rho_{eff}$ and $p_{eff}$ are the effective energy density and pressure, respectively. In the following subsections, we will derive the energy conditions for the two models being analyzed. It is important to emphasize that our consideration of the effective quantities stems from their ability to account for the modifications to the gravitational theory in the Weyl-type $f(Q,T)$ function.

\begin{widetext}
\subsection{$f(Q,T)=\protect\alpha Q+\frac{\protect\beta }{6\protect\kappa ^2%
}T$}

Using Eqs. (\ref{rho1}) and (\ref{p1}), we get the energy conditions of the
form 
\begin{eqnarray}
\rho &=& -\frac{3\left( H^{2}\left( 12(2\alpha \beta +3\alpha
+\beta )+(2\beta +3)M^{2}\right) +4\beta \dot{H}\right) }{2\left( 2\beta
^{2}+9\beta +9\right) }\geq 0, \\
\rho +p &=& -\frac{3\left( H^{2}\left( 12(\alpha +1)+M^{2}\right)
+4\dot{H}\right) }{\beta +3}\geq 0, \\
\rho -p &=& \frac{12\left( 3H^{2}+\dot{%
H}\right) }{2\beta +3}\geq 0, \\
\rho +3p &=& -\frac{6\left( H^{2}\left( 6(4\alpha \beta +6\alpha
+5\beta +9)+(2\beta +3)M^{2}\right) +2(5\beta +9)\dot{H}\right) }{2\beta
^{2}+9\beta +9}\geq 0.
\end{eqnarray}

\subsection{$f(Q,T)=\frac{\protect\gamma}{6H_0^2\protect\kappa ^2}QT$}

Using Eqs. (\ref{rho2}) and (\ref{p2}), we get the energy conditions of the
form%
\begin{eqnarray}
\rho &=& \frac{H^2 H_{0}^2 \left(2 \gamma  H^2 \left(M^2+60\right)+40 \gamma  \dot{H}-H_{0}^2 M^2\right)}{2 \left(8 \gamma ^2 H^4+4 \gamma  H^2 H_{0}^2+H_{0}^4\right)}\geq 0, \\
\rho +p &=& \frac{16\gamma H^{2}H_{0}^{2}\left( 3H^{2}+\overset{.}%
{H}\right) -H_{0}^{4}\left( H^{2}\left( M^{2}+12\right) +4\dot{H}\right) }{%
8\gamma ^{2}H^{4}+4\gamma H^{2}H_{0}^{2}+H_{0}^{4}}\geq 0, \\
\rho -p &=& \frac{2H_{0}^{2}\left(
\gamma H^{4}\left( M^{2}+36\right) +2\dot{H}\left( 6\gamma
H^{2}+H_{0}^{2}\right) +6H^{2}H_{0}^{2}\right) }{8\gamma ^{2}H^{4}+4\gamma
H^{2}H_{0}^{2}+H_{0}^{4}}\geq 0, \\
\rho +3p &=& -\frac{2H_{0}^{2}\left( \gamma H^{4}\left(
M^{2}-12\right) +\dot{H}\left( 6H_{0}^{2}-4\gamma H^{2}\right)
+H^{2}H_{0}^{2}\left( M^{2}+18\right) \right) }{8\gamma ^{2}H^{4}+4\gamma
H^{2}H_{0}^{2}+H_{0}^{4}}\geq 0.
\end{eqnarray}
\end{widetext}

\section{Constraints from Observational Data}
\label{sec5}

The given system of Eqs. (\ref{F1}) and (\ref{F2}) represents a set of two equations with three unknowns: $p$, $\rho$, and $H$. To analyze the energy conditions and further investigate the system, an additional physically meaningful condition needs to be imposed. In this particular study, we choose to consider the deceleration parameter as a means to establish this condition. In this analysis, we will consider the parametrization form of the deceleration parameter (DP) as,
\begin{equation}
    q(z)=1-\frac{2}{1+\chi  (1+z)^4},
    \label{DP}
\end{equation}
where $z$ represents the redshift and $\chi$ is a parameter that determines the rate of deceleration.

The parameterization method of the deceleration parameter has been widely used in cosmological studies as it provides a flexible and convenient way to explore different expansion scenarios of the Universe \cite{P1,P2,P3,P4,P5,P6}. Here are some motivations behind this parametrization form given by Eq. (\ref{DP}):
\begin{itemize}
    \item \textbf{Redshift dependence} \cite{Riess,Perlmutter}: The deceleration parameter, which characterizes the rate of expansion of the Universe, is expected to evolve with cosmic time. The parametrization form incorporates the redshift $z$ as a variable, allowing us to study the behavior of the deceleration parameter at different cosmic epochs.
    \item \textbf{Model-independent approach} \cite{P7,P8}: The parametrization form does not rely on specific cosmological models or assumptions. Instead, it provides a phenomenological description that can be applied to various theoretical frameworks, including both DE and modified gravity models.
    \item \textbf{Flexibility in cosmic acceleration}: By adjusting the parameter $\chi$, the parametrization allows for a wide range of possibilities for cosmic acceleration. It enables the exploration of both accelerating and decelerating phases of the Universe, as well as transitions between them.
    \item \textbf{Comparison with observational data}: The parametrization form can be constrained using observational data, such as supernovae type Ia, baryon acoustic oscillations, and cosmic microwave background observations. By fitting the parametrization to the data, one can test different cosmological scenarios and evaluate the consistency with observations.
\end{itemize}

In addition, below are the values of the parametrized deceleration parameter for different cosmic epochs:
\begin{itemize}
    \item \textbf{Past Universe (Early Universe)}: At very early times, when $z \gg 1$, the deceleration parameter approaches $q(z) \approx 1$. This corresponds to a decelerating Universe dominated by matter and radiation.
    \item \textbf{Present Universe}: For the present epoch, with $z \approx 0$, the deceleration parameter is given by $q(z) = 1-\frac{2}{1+\chi}$. Depending on the value of $\chi$, the deceleration parameter can take different values. For $\chi = 0$, we have $q(z) = 1$, indicating a matter-dominated Universe with no cosmic acceleration. For $\chi > 0$, the deceleration parameter becoming smaller than 1 does not necessarily imply an accelerating Universe. So, the specific value of $\chi$ determines the strength of the acceleration, with larger values indicating a more rapid expansion.
    \item \textbf{Future Universe}: As we move towards the future, with increasing redshift $z < 0$, the deceleration parameter evolves depending on the chosen value of $\chi$. If $\chi = 0$, the deceleration parameter asymptotically approaches $q(z) \approx -1$, indicating a Universe undergoing constant acceleration. This behavior implies an exponential expansion known as the de Sitter phase. Also, for distant future, $z\approx=-1$ implying $q(z)\approx -1$.
\end{itemize}

It's important to note that the specific values of the deceleration parameter depend on the chosen value of $\chi$. Different values of $\chi$ lead to different cosmological scenarios, ranging from deceleration to acceleration in various cosmic epochs.

The relationship between the deceleration parameter and the Hubble parameter can be expressed by the following equation:
\begin{equation}
    H(z)=H_{0} exp\left(\int_0^z \frac{1+q(z)}{(1+z)} \, dz\right).
    \label{hz}
\end{equation}

By substituting Eq. (\ref{DP}) into Eq. (\ref{hz}), we obtain the expression for $H(z)$ as follows:
\begin{equation}
    H(z)=H_{0}\left[A(1+z)^4+B\right]^\frac{1}{2},
\end{equation}
where $H_{0}$ is the present value of $H(z)$, $A=\frac{\chi}{1+\chi}$ and $B=\frac{1}{1+\chi}$. Further, the derivative of the Hubble parameter with respect to time can be expressed as
\begin{equation}
\dot{H}=\frac{dH}{dt}=-\left(1+z\right)H(z)\frac{dH}{dz}.
\end{equation}

Next, we assess the validity of the parametrization by examining its consistency with recent observational data, specifically the observational Hubble data, and Type Ia supernovae (SNe Ia):
\begin{itemize}
\item \textbf{Hubble dataset}: We consider 31 Hubble data points obtained using the differential age (DA) approach \cite{Yu/2018, Moresco/2015, Sharov/2018}.
\item \textbf{SNe Ia dataset}: We also include 1048 SNe Ia luminosity distance estimates from the Pan-STARSS 1 (PS1) Medium Deep Survey, the Low-z, SDSS, SNLS, and HST missions in the Pantheon sample \cite{Scolnic/2018, Chang/2019}.
\end{itemize}

To analyze the cosmological observational data, we adopt the Markov Chain Monte Carlo (MCMC) sampling technique. Our approach expands on existing studies by incorporating an extensive range of data and applying more stringent priors on the model parameters. Specifically, we focus on the parameter space $\theta_{s}=(H_{0},\chi)$ and employ the \textit{emcee} library \cite{Mackey/2013} for parallelized MCMC sampling. Our analysis utilizes 100 walkers and 1000 steps to obtain reliable results. By combining the information from the two types of data (Hubble and Pantheon), we extract valuable insights into the model parameters and their uncertainties.

We define the $\Tilde{\chi}^2$ function for the joint Hubble+Pantheon data as
\begin{equation}
\Tilde{\chi}^{2}_{joint}=\Tilde{\chi}^{2}_{Hubble}+\Tilde{\chi}^{2}_{Pantheon},
\end{equation}
where
\begin{equation}
\Tilde{\chi}^{2}_{Hubble} = \sum_{i=1}^{31} \frac{\left[H(\theta_{s}, z_{i})-
H_{obs}(z_{i})\right]^2}{\sigma(z_{i})^2},
\end{equation}
and
\begin{equation}
\Tilde{\chi}^{2}_{Pantheon} =\sum_{i,j=1} ^{1048} \Delta \mu_{i} \left(
C_{Pantheon}^{-1}\right)_{ij} \Delta \mu_{j}
\end{equation}

Here, the variables are defined as follows: $H(z_{i})$ represents the theoretical value of the Hubble parameter for a specific model at different redshifts $z_{i}$, $H_{obs}(z_{i})$ is the observed value of the Hubble parameter, $\sigma(z_{i})$ represents the observational error, $\Delta \mu_{i}=\mu_{\rm th}-\mu_{\rm obs}$ represents the difference between the theoretical and observed distance modulus, and $C_{Pantheon}^{-1}$ is the inverse of the covariance matrix of the Pantheon sample. For more detailed information, please refer to the following Refs. \cite{Q6,Q8,Weyl3}.

\begin{figure}[h]
\centerline{\includegraphics[scale=0.6]{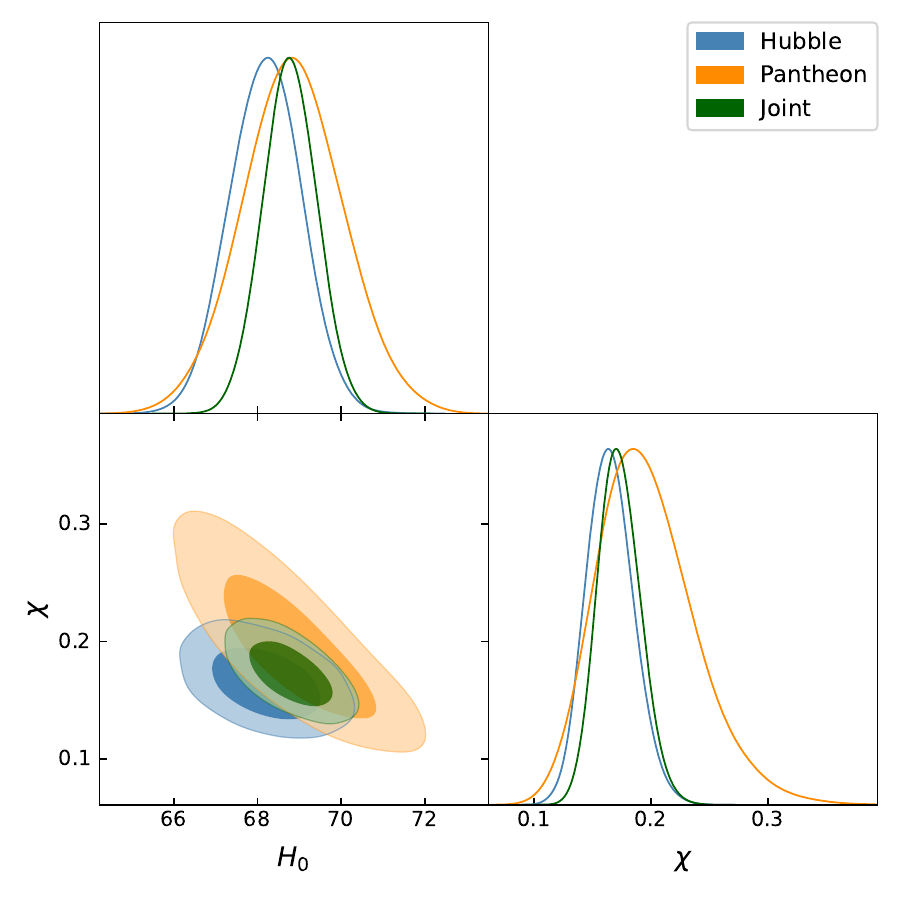}}
\caption{The likelihood contours of the $1-\sigma$ and $2-\sigma$ confidence levels of the model parameters $H_{0}$ $(km/s/Mpc)$ and $\chi$ using the Hubble and Pantheon samples.}
\label{Combine}
\end{figure}

The observational constraints on the model parameters are obtained by minimizing the corresponding $\Tilde{\chi}^2$ using the MCMC method. The results of this analysis, including the best-fit values and uncertainties for the model parameters, are summarized in Tab. \ref{tab}. Further, the $1-\sigma$ and $2-\sigma$ likelihood contours on the model parameters are shown in Fig. \ref{Combine}, providing a graphical representation of the allowed parameter regions. In Figs. \ref{ErrorHubble} and \ref{ErrorSNe}, the error bar fit for the considered model is shown, along with the $\Lambda$CDM model with $\Omega_{m0} = 0.315$, and $H_0 = 67.4$ $km/s/Mpc$ \cite{Planck}. These figures illustrate the goodness of fit for the model and provide a visual comparison with the standard $\Lambda$CDM model. While our analysis reveals that the values of $H_0$ for the Hubble dataset, Pantheon, and the combined dataset are close to those of the $\Lambda$CDM model, suggesting consistency with the standard model, it is important to note that the determination of which model is superior is not solely dependent on $H_0$. The superiority of a cosmological model is determined by its ability to provide a more complete and accurate description of the universe's evolution, including its ability to explain a wide range of observational data beyond just the value of $H_0$. In this context, while the values of $H_0$ are comparable between our model and the $\Lambda$CDM model, our model offers additional advantages, such as a more flexible description of the cosmological evolution, better fit to observational data, and theoretical consistency.

\begin{widetext}

\begin{table*}[!htbp]
\begin{center}
\begin{tabular}{l c c c c c c}
\hline\hline 
$datasets$              & $H_{0}$ ($km/s/Mpc$) & $\chi$ & $q_{0}$ & $z_{tr}$ & $\omega_{0}$ (Model 1) & $\omega_{0}$ (Model 2) \\
\hline
$Priors$   & $(60,80)$  & $(-10,10)$  & $-$ & $-$ & $-$ & $-$\\
$Hubble$ & $68.2_{-1.7}^{+1.7}$  & $0.165_{-0.037}^{+0.042}$  & $-0.72^{+0.06}_{-0.06}$ & $0.57^{+0.09}_{-0.08}$ & $-0.80^{+0.05}_{-0.04}$ & $-0.89^{+0.01}_{-0.01}$\\
$Pantheon$   & $68.9_{-2.4}^{+2.4}$  & $0.197_{-0.078}^{+0.085}$  & $-0.67^{+0.11}_{-0.1}$ & $0.50^{+0.13}_{-0.12}$ & $-0.76^{+0.08}_{-0.07}$ & $-0.89^{+0.01}_{-0.01}$\\
$Joint$   & $68.8_{-1.3}^{+1.3}$  & $0.173_{-0.034}^{+0.038}$  & $-0.71^{+0.06}_{-0.05}$ & $0.55^{+0.07}_{-0.07}$ & $-0.79^{+0.04}_{-0.04}$ & $-0.89^{+0.01}_{-0.01}$\\

\hline\hline
\end{tabular}
\caption{The results of MCMC on cosmological parameters using the Hubble and Pantheon samples.}
\label{tab}
\end{center}
\end{table*}

\begin{figure}[!htb]
   \begin{minipage}{1.0\textwidth}
     \centering
     \includegraphics[width=.7\linewidth]{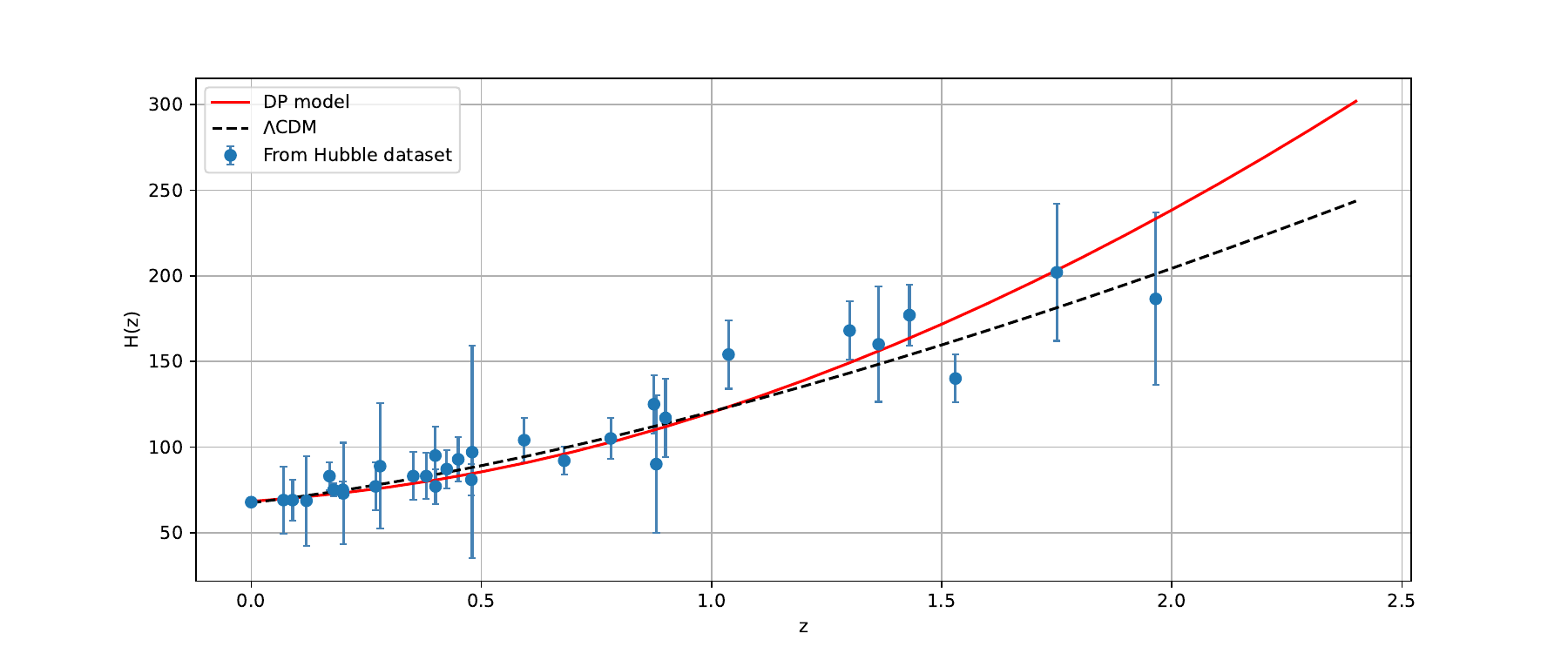}
     \caption{The evolution of Hubble parameter $H(z)$ as a function of redshift $z$: Comparison between $\Lambda$CDM model (black dashed line) and DP model (red line) with error bars (steel-blue dots)}\label{ErrorHubble}
   \end{minipage}\hfill
   \begin{minipage}{1.0\textwidth}
     \centering
     \includegraphics[width=.7\linewidth]{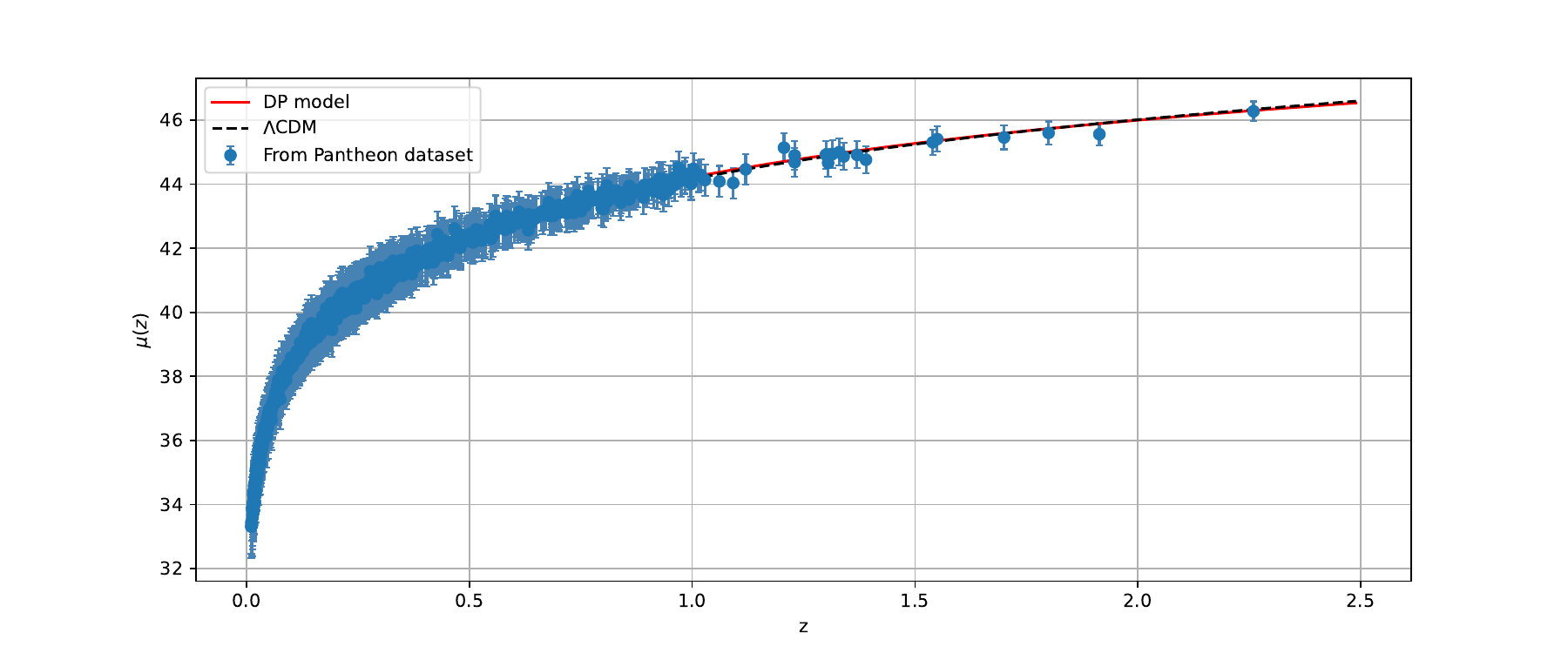}
     \caption{The evolution of distance modulus $\mu(z)$ as a function of redshift $z$: Comparison between $\Lambda$CDM model (black dashed line) and DP model (red line) with error bars (steel-blue dots)}\label{ErrorSNe}
   \end{minipage}
\end{figure}

\end{widetext}

\section{Results and Discussion}
\label{sec6}

\begin{figure}[!htb]
     \centering
     \includegraphics[width=.7\linewidth]{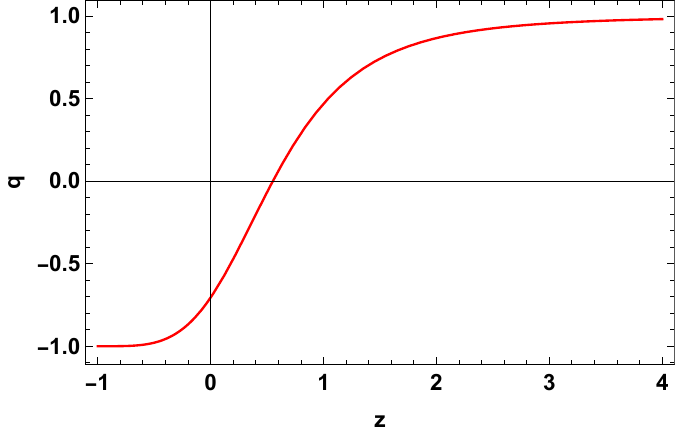}
     \caption{The evolution of the deceleration parameter $q$ as a function of redshift $z$ for the constrained values of model parameters from the joint data analysis.}\label{F_q}
\end{figure}

The behavior of the deceleration parameter $q$ provides valuable insights into the dynamics of cosmic acceleration throughout the evolution of the Universe. Fig. \ref{F_q} showcases the intriguing findings of our analysis for the constrained values of the model parameters obtained from the joint Hubble+Pantheon data analysis, revealing the transition of $q$ from a decelerated phase $q>0$ to an accelerated phase $q<0$ at a specific redshift $z_{tr}$ (i.e. $q=0$). This transition signifies a significant shift in the dominant cosmic forces driving the expansion. The observed value of $z_{tr}=0.55^{+0.07}_{-0.07}$, as reported by recent observations \cite{Dinda,Almada}, further validates the credibility of our results. The value of $z_{tr}$ aligns with the current understanding of cosmic acceleration and is consistent with various independent measurements. Furthermore, for the model parameters constrained by the Hubble+Pantheon dataset, the present value of the deceleration parameter is $q_{0}=-0.71^{+0.06}_{-0.05}$ \cite{Wu,Cunha}. Therefore, the detailed investigation of $q$ and its corresponding redshift $z_{tr}$ is crucial for comprehending the underlying physics governing the expansion of the Universe. It provides important constraints on cosmological models and sheds light on the nature of DE or modifications to gravity at different cosmic epochs. Our findings demonstrate the capability of the employed model to capture the complex dynamics of the Universe, incorporating both decelerated and accelerated phases of expansion. This comprehensive understanding of cosmic evolution enhances our knowledge of the underlying cosmological mechanisms and brings us closer to unraveling the mysteries of the Universe's past, present, and future.

The EoS parameter $\omega$ also provides valuable insights into the nature of cosmic expansion by characterizing the relationship between energy density $\rho$ and pressure $p$. By examining the behavior of $\omega$, we can classify the different phases of the Universe's expansion. For instance, when $\omega=1$, it represents a stiff fluid, indicating an extreme scenario where pressure is equal to energy density. On the other hand, the matter-dominated phase is characterized by $\omega=0$, signifying a scenario where pressure is negligible compared to energy density. In the radiation-dominated phase, we observe $\omega=\frac{1}{3}$, indicating a state where radiation dominates the cosmic dynamics. Furthermore, when $\omega$ lies within the range $-1 <\omega<-\frac{1}{3}$, it represents the quintessence phase, where a scalar field with a time-evolving EoS drives the accelerated expansion \cite{Quin1, Quin2}. The special case of $\omega=-1$ corresponds to the cosmological constant, commonly referred to as the $\Lambda$CDM model \cite{weinberg/1989}, which accurately describes the observed acceleration of the Universe. In certain scenarios, we may encounter a phase known as the phantom era, characterized by $\omega<-1$ \cite{phantom1,phantom2}. During this phase, the pressure becomes more negative than the energy density, leading to exotic phenomena and potential instabilities in cosmic evolution. It is worth noting that for an accelerating Universe, $\omega$ must satisfy the condition $\omega<-1/3$. This condition is equivalent to a violation of the SEC. The SEC is one of the energy conditions in GR that imposes certain constraints on the behavior of energy and matter in the Universe. According to the SEC, the sum of the energy density and three times the pressure should always be non-negative, which can be expressed as $\rho+3p\geq0$. This condition ensures that the gravitational effects of matter and energy are attractive and prevents the existence of exotic forms of matter with repulsive gravity. However, in the case of an accelerating Universe, where the expansion rate is increasing over time, the violation of the strong energy condition ($\rho+3p<0$ or $\omega<-1/3$) becomes necessary \cite{Barcelo}. 

\subsection{$f(Q,T)=\protect\alpha Q+\frac{\protect\beta }{6\protect\kappa ^2%
}T$}

In order to ensure a physically meaningful and observationally consistent cosmological model, specific values for the model parameters need to be chosen. In this study, we have selected the values $\alpha = -1.07$ and $\beta = 0.5$ for this purpose. These parameter values have been chosen to satisfy two important criteria: ensuring the positivity of the energy density and aligning the EoS parameter with observational constraints. 

Fig. \ref{F_rho1} illustrates the relationship between the density parameter and redshift. One can observe that the density parameter is consistently positive for all redshift values. Moreover, at redshift $z=0$, the density parameter is strictly positive and as the redshift increases, the energy density also increases. This behavior confirms that the WEC is satisfied in this model. It indicates that the Universe's energy content remains non-negative and increases as the Universe expands. The behavior of the EoS parameter in Fig. \ref{F_EoS1} clearly indicates that $-1<\omega<-1/3$, suggesting the presence of quintessence DE and implying an accelerating phase of the Universe. It is noteworthy that the current value of the EoS parameter for this case is $\omega_{0}=-0.79^{+0.04}_{-0.04}$ \cite{EoS1,EoS2,Gruber,Basilakos} for the joint Hubble+Pantheon data.

The energy conditions for this particular scenario are depicted in Figs. \ref{F_SEC1}, \ref{F_NEC1} and \ref{F_DEC1}. 

The violation of the SEC on a cosmological scale is an intriguing aspect of modern cosmology. It suggests the presence of exotic forms of energy or modifications to the laws of gravity at cosmic scales. In light of the SEC's significance, we examined the acceptable ranges of the model parameter $\beta$, as depicted in Fig. \ref{F_SEC1}. The parameter $\alpha$ is fixed to be $-1.07$, while $\beta$ ranges from $0.5$ to $3.0$. The variation in $\beta$ leads to changes in the behavior of the SEC. For values of $\beta<0$, the SEC exhibits some positive behavior. We focused on the range of $\beta<0$ where the violation of the SEC is more pronounced. Furthermore, a negative value of $\omega$ implies $\rho + 3p < 0$, indicating a violation of the SEC at the present epoch. The results shown in Figs. \ref{F_NEC1} and \ref{F_DEC1} demonstrate that the NEC and DEC are satisfied for this particular scenario. These conclusions are supported by the behavior of the energy density shown in Fig. \ref{F_rho1}. Thus, the validation of the NEC and the energy density together confirm the fulfillment of the WEC.

\begin{widetext}

\begin{figure}[!htb]
   \begin{minipage}{0.48\textwidth}
     \centering
     \includegraphics[width=.7\linewidth]{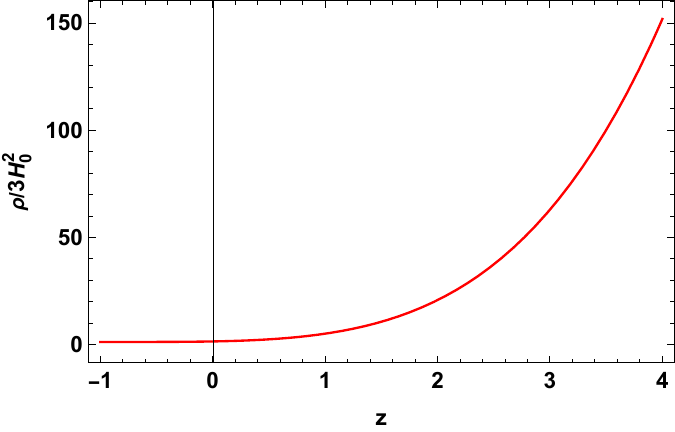}
     \caption{The evolution of the density parameter $\rho_{1}$ as a function of redshift $z$ (linear model).}\label{F_rho1}
   \end{minipage}\hfill
   \begin{minipage}{0.48\textwidth}
     \centering
     \includegraphics[width=.7\linewidth]{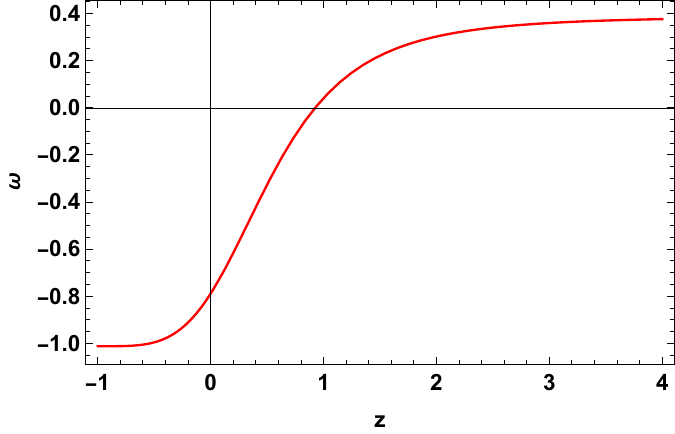}
     \caption{The evolution of the EoS parameter $\omega_{1}$ as a function of redshift $z$ (linear model).}\label{F_EoS1}
   \end{minipage}
\end{figure}

\begin{figure}
     \centering
     \begin{subfigure}[b]{0.33\textwidth}
         \centering
         \includegraphics[width=\textwidth]{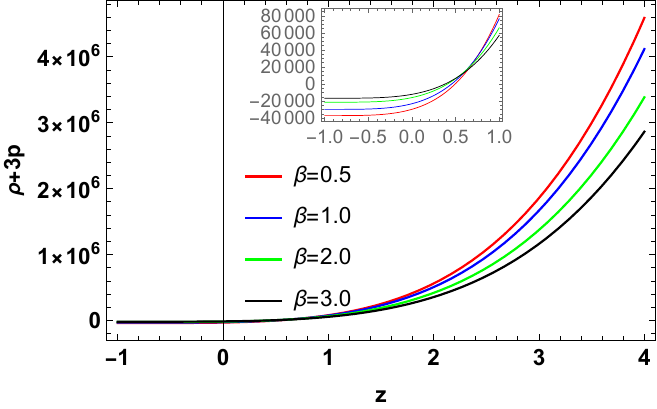}
         \caption{$SEC:\rho+3p$}
         \label{F_SEC1}
     \end{subfigure}
     \hfill
     \begin{subfigure}[b]{0.33\textwidth}
         \centering
         \includegraphics[width=\textwidth]{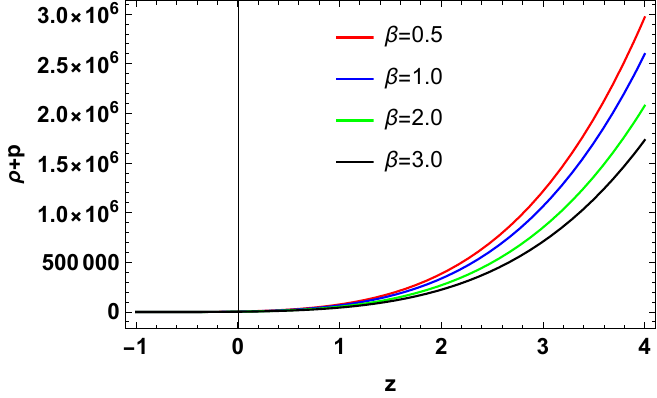}
         \caption{$NEC:\rho+p$}
         \label{F_NEC1}
     \end{subfigure}
     \hfill
     \begin{subfigure}[b]{0.33\textwidth}
         \centering
         \includegraphics[width=\textwidth]{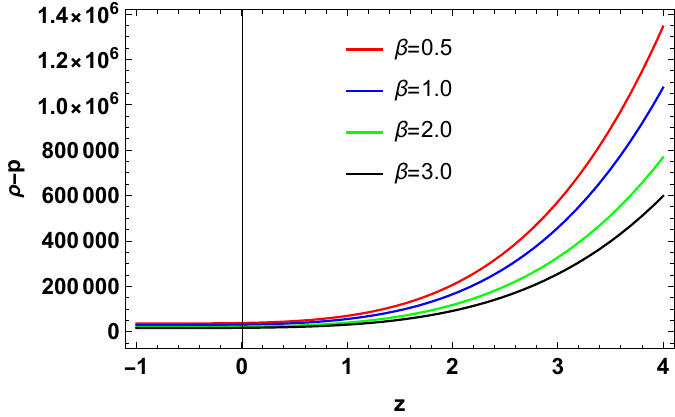}
         \caption{$DEC:\rho-p$}
         \label{F_DEC1}
     \end{subfigure}
        \caption{The evolution of the energy conditions as a function of redshift $z$ (linear model).}
        \label{F_ECs1}
\end{figure}
\end{widetext}

\subsection{$f(Q,T)=\frac{\protect\gamma}{6H_0^2\protect\kappa ^2}QT$}

The density and EoS parameters corresponding to this scenario are illustrated below.

Fig. \ref{F_rho2} presents the relationship between the density parameter and redshift, considering the parameter values $\gamma=0.315$. It is evident that the density parameter is consistently positive for all redshift values. At $z = 0$, the density parameter is strictly positive and it grows as the redshift increases. This behavior is in line with our understanding of an expanding Universe where the energy density of cosmic components, such as DE, decreases with time. The behavior of the EoS parameter in Fig. \ref{F_EoS2} exhibits a similar pattern to the previous model. It is found that the EoS parameter lies within the range $-1<\omega<-1/3$, which implies the presence of quintessence DE and an accelerating phase of the Universe. Specifically, for this scenario, the current value of the EoS parameter is determined to be $\omega_{0}=-0.89^{+0.01}_{-0.01}$ \cite{EoS1,EoS2,Gruber,Basilakos} (for the joint data). This value indicates that the DE component in the Universe is consistent with quintessence, driving the accelerated expansion observed in cosmological observations.

In Fig. \ref{F_SEC2}, we show the behavior of the SEC while varying the parameter $\gamma$ in the small range of $0.315$ to $0.340$. It is observed that with the variation of $\gamma$, the SEC is violated at the present epoch. As $\gamma$ moves with a small change, there is a noticeable change in the SEC behavior, and it exhibits a more pronounced violation within the mentioned range. The negative behavior of the SEC indicates the accelerated expansion of the Universe, consistent with the presence of DE. On the other hand, the NEC and the DEC are not violated, as their behaviors remain positive throughout (refer to Figs. \ref{F_NEC2} and \ref{F_DEC2}). The satisfaction of NEC along with the positive behavior of the density parameter confirms the validity of the WEC.

\begin{widetext}

\begin{figure}[!htb]
   \begin{minipage}{0.48\textwidth}
     \centering
     \includegraphics[width=.7\linewidth]{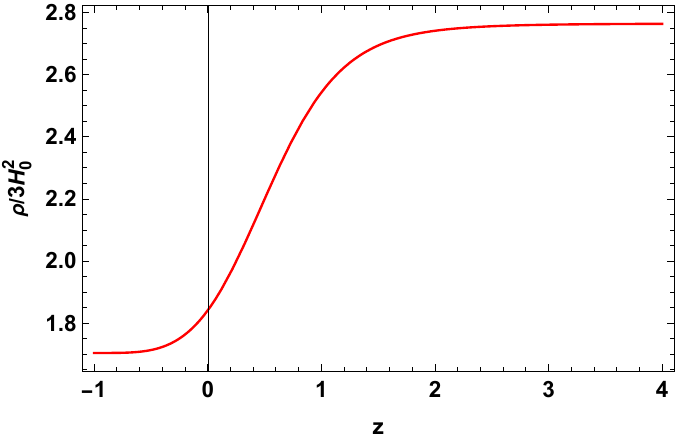}
     \caption{The evolution of the density parameter $\rho_{2}$ as a function of redshift $z$ (coupling model).}\label{F_rho2}
   \end{minipage}\hfill
   \begin{minipage}{0.48\textwidth}
     \centering
     \includegraphics[width=.7\linewidth]{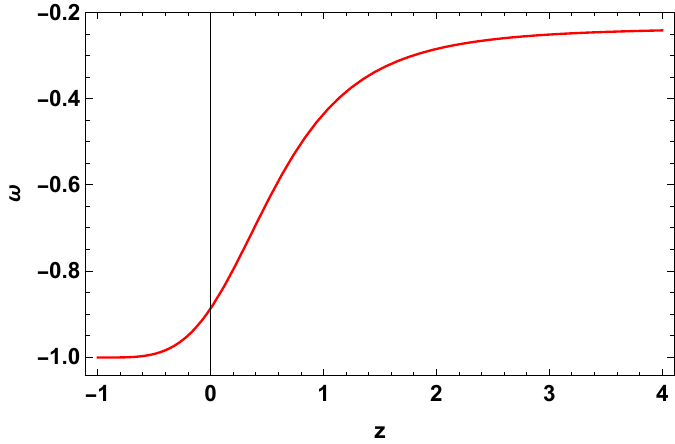}
     \caption{The evolution of the EoS parameter $\omega_{2}$ as a function of redshift $z$ (coupling model).}\label{F_EoS2}
   \end{minipage}
\end{figure}
    
\begin{figure}
     \centering
     \begin{subfigure}[b]{0.33\textwidth}
         \centering
         \includegraphics[width=\textwidth]{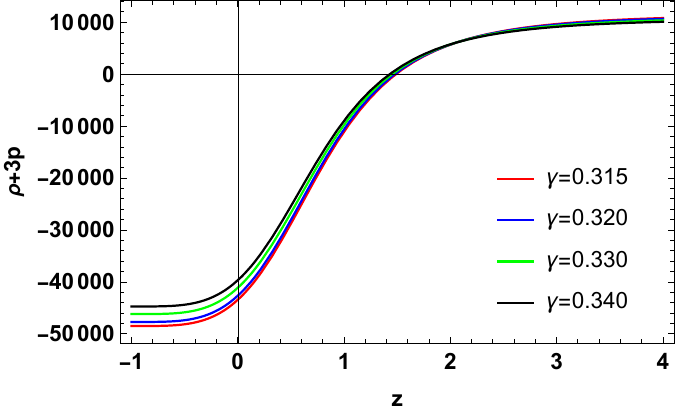}
         \caption{$SEC:\rho+3p$}
         \label{F_SEC2}
     \end{subfigure}
     \hfill
     \begin{subfigure}[b]{0.33\textwidth}
         \centering
         \includegraphics[width=\textwidth]{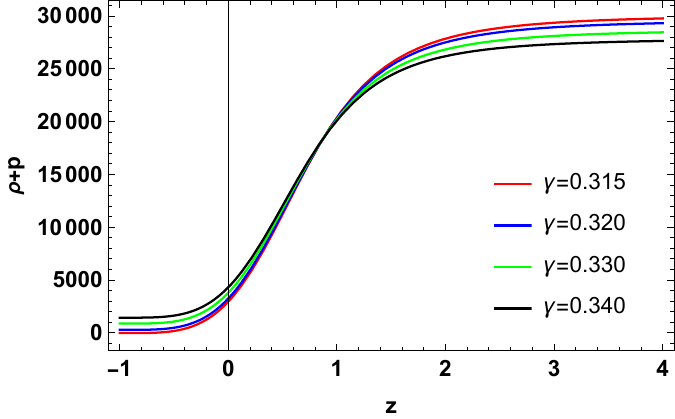}
         \caption{$NEC:\rho+p$}
         \label{F_NEC2}
     \end{subfigure}
     \hfill
     \begin{subfigure}[b]{0.33\textwidth}
         \centering
         \includegraphics[width=\textwidth]{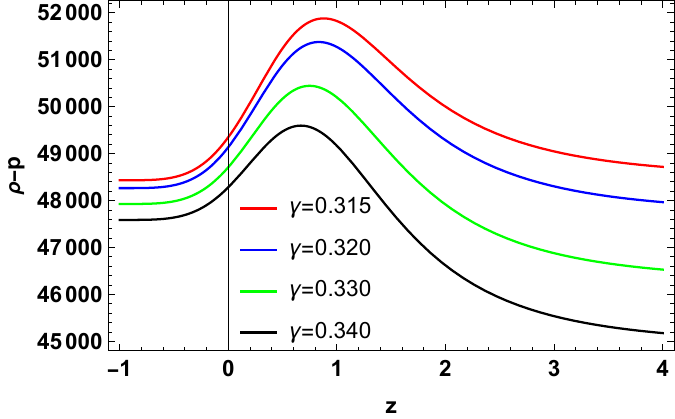}
         \caption{$DEC:\rho-p$}
         \label{F_DEC2}
     \end{subfigure}
        \caption{The evolution of the energy conditions as a function of redshift $z$ (coupling model).}
        \label{F_ECs2}
\end{figure}
\end{widetext}

\section{Conclusion}
\label{sec7}

Recent observational studies provide compelling evidence for the accelerated expansion of the Universe. This accelerated expansion is attributed to the presence of a high negative pressure component called DE, which satisfies the condition $\rho+3p<0$ or $\omega<-1/3$. This observation has motivated further investigations into the nature of gravity and led to the development of modified theories of gravity, which aim to provide a geometrical extension to GR. These modified theories of gravity offer alternative explanations for the observed cosmic acceleration and seek to reconcile them with our understanding of fundamental physics.

In this paper, we have investigated the cosmological implications and constraints of Weyl-type $f(Q, T)$ gravity. This theory introduces a coupling between the non-metricity $Q$ and the trace $T$ of the energy-momentum tensor, using the principles of proper Weyl geometry. In this geometry, the scalar non-metricity $Q$, which characterizes the deviations from Riemannian geometry, is expressed in its standard Weyl form and is determined by a vector field $w_{\alpha }$. To study the implications of this theory, we have proposed a deceleration parameter with a single unknown parameter $\chi$. By solving the field equations derived from Weyl-type $f(Q, T)$ gravity, we have studied the behavior of the energy conditions within this framework. The energy conditions provide important constraints on the physical viability of a theory, and by examining their satisfaction or violation, we have gained valuable insights into the nature of the gravitational theory and its compatibility with observational data. Especially, we have employed two well-motivated forms of the function $f(Q, T)$ to investigate the gravitational theory: (i) the linear model represented by $f(Q, T) = \alpha Q + \frac{\beta}{6\kappa^2} T$, and (ii) the coupling model represented by $f(Q, T) = \frac{\gamma}{6H_0^2 \kappa^2} QT$. We further investigated the behavior of the deceleration parameter with respect to redshift, considering the values $H_{0}=68.8_{-1.3}^{+1.3}$  and $\chi=0.173_{-0.034}^{+0.038}$, which were obtained from the joint Hubble+Pantheon dataset. The deceleration parameter $q$ exhibits a transition from an early decelerating phase to the current accelerated expansion of the Universe at a redshift of $z_{tr}=0.55^{+0.07}_{-0.07}$, consistent with the literature values. The present value of the deceleration parameter $q_{0}$ is determined to be $-0.71^{+0.06}_{-0.05}$, aligning with previous findings. It is very important to note that the parameters of the deceleration parameter have been constrained using different datasets (Hubble and Pantheon), as presented in Tab. \ref{tab} and Fig. \ref{Combine}. By observing the convergence of these parameter values, our focus has shifted towards analyzing the joint dataset to derive more robust and reliable results.

In Sec. \ref{sec6}, we have investigated the behavior of density, EoS parameter, and energy conditions for the two models of Weyl-type $f(Q,T)$ gravity. The density parameter in both models exhibits positive behavior, indicating a non-negative energy content of the Universe. Furthermore, the EoS parameter analysis reveals that it lies within the range of $-1<\omega<-1/3$, signifying the presence of quintessence DE and implying an accelerating phase of the Universe \cite{Quin1,Quin2}. The values of the model parameters $\alpha$, $\beta$, and $\gamma$ are further used to investigate the behavior of various energy conditions in relation to the EoS parameter. In both the models presented in Sec. \ref{sec6}, the energy conditions, namely the NEC, WEC, and DEC, are satisfied. This is observed through the analysis of the corresponding figures, such as Figs. \ref{F_ECs1} and \ref{F_ECs2}. However, it should be noted that the SEC is violated in these models. This violation of SEC is indicative of the accelerated expansion of the Universe and is consistent with previous studies and observations.

In this study, we have focused on two specific cases of Weyl-type $f(Q,T)$ gravity, namely the linear and coupling models. These models have provided valuable insights into the behavior of energy conditions and the dynamics of the Universe. In the Weyl-type $f(Q,T)$ gravity theory, the cosmological evolution's nature is significantly influenced by the values of the model parameters and the functional form of $f(Q,T)$. Our analysis of specific models and a range of cosmological parameters reveals a fundamental result: the universe initiated its recent evolution in a decelerating phase before transitioning into an accelerating phase. The values of the model parameters allow for the construction of a wide range of cosmological scenarios, encompassing the evolution of the universe through the radiation era ($\omega=\frac{1}{3}$), matter era ($\omega=0$), quintessence era ($-1<\omega<-\frac{1}{3}$), and $\Lambda$CDM model ($\omega=-1$). Remarkably, the linear model predicted three phases of the universe (see Fig. \ref{F_EoS1}), similar to the $\Lambda$CDM model (radiation, matter, and acceleration phases), whereas the coupling model predicts only the acceleration phase (see Fig. \ref{F_EoS2}). In contrast to the standard $\Lambda$CDM model, our models exhibit quintessence-like behavior akin to a DE model. Despite deviating from the $\Lambda$CDM paradigm in certain aspects, the variant of the Weyl-type $f(Q,T)$ modified gravity theory considered here remains capable of explaining current observational data on cosmological parameters. It provides a compelling and internally consistent explanation for the accelerating expansion of the Universe. However, it is important to note that our analysis is not exhaustive, and there are still other possible models to explore. One such example is the exponential case i.e. $f(Q,T)=\eta H_0^2e^{\frac{\mu}{6H_0^2} Q}+\frac{\nu}{6\kappa ^2} T$, which was mentioned as the third model in the original paper \cite{Weyl1}. Future investigations can delve into studying this particular model and its implications for the energy conditions and the evolution of the Universe.

Further, in light of the limitations of our models, a key question arises regarding the feasibility of directly solving the field equations in Weyl-type $f(Q,T)$ gravity, as opposed to resorting to the parametrization method (such as Eq. (\ref{DP})) \cite{Q6,Q7,Q8}. It is worth noting that obtaining exact solutions to these field equations is highly challenging due to their complexity. However, despite the current difficulties, future research efforts may aim to tackle this challenge and explore the possibility of obtaining direct solutions.

\section*{Acknowledgments}
The authors extend their appreciation to the Deputyship for Research \& Innovation, Ministry of Education in Saudi Arabia for funding this research through project number IFP-IMSIU-2023122. The authors also appreciate the Deanship of Scientific Research at Imam Mohammad Ibn Saud Islamic University (IMSIU) for supporting and supervising this project.

\textbf{Data availability} There are no new data associated with this
article.

\textbf{Conflict of interest} The authors declare that they have no competing interests.

\end{document}